\documentclass[aps,prl,showpacs,preprintnumbers,amsmath,amssymb,12pt]{revtex4}
\usepackage{dcolumn}
\usepackage[colorlinks,dvipdfm]{hyperref}
\usepackage{epsf}

\newcommand{\be}{\begin{equation}}
\newcommand{\ee}{\end{equation}}
\newcommand{\ba}{\begin{array}}
\newcommand{\bqa}{\begin{eqnarray}}
\newcommand{\eqa}{\end{eqnarray}}
\newcommand{\cO}{{\cal O}}

\begin{document}

\begin{flushright}
CAFPE-133/09\\
UG-FT-263/09\\
\end{flushright}

\title{\bf Positivity constraints on  LECs of $\chi$PT lagrangian at $\cO(p^6)$ level}

\author{Zhi-Hui~Guo$^1$, Ou Zhang$^2$ and H.~Q.~Zheng$^{2}$}

\affiliation{ 1: CAFPE and Departamento de Fisica Teorica y del
Cosmos, Universidad de Granada,
 Campus de Fuente Nueva, E-18002 Granada, Spain.
\\
2:  Department of Physics, Peking University,Beijing 100871,
P.~R.~China.}

\begin{abstract}
 Positivity constraints on the LECs of $\cO(p^6)$
$\chi$PT lagrangian are discussed. We demonstrate that the
constraints are automatically satisfied inside the Mandelstam
triangle for $\pi\pi$ scatterings, when $N_C$ is large. Numerical
tests are made in the $N_C=3$ case, and it is found that these
constraints are also  well respected.
\end{abstract}

\vskip .5cm


\date{\today}
\maketitle

 The concept of effective field theory plays one of the central role in
 modern particle physics. A well-known example is
  chiral perturbation theory ($\chi$PT) which is crucial
  in the study of  low energy hadron physics~\cite{chPT}. It describes the
interaction  between pseudo-Goldstone bosons of QCD and its
lagrangian is constructed based on the expansion of the external
momentum and the mass of the pseudo-Goldstone bosons. In the leading
order, only two parameters are involved: the pion decay constant and
the pion mass (In this letter, we only focus on $SU(2)$ $\chi$PT).
When stepping into higher orders, there appears a number of low
energy constants (LECs), which are free parameters of the chiral
lagrangian and are not fixed by chiral symmetry requirement.
Nevertheless, it is possible to obtain certain constraints on these
LECs, using general properties that a quantum field theory has to
obey, like analyticity, unitarity and positivity. Efforts have been
made in the literature to understand these LECs along this line. In
Ref.~\cite{dispersive} positivity constraints on LECs are carefully
studied at $\cO(p^4)$ level, and it is found that these constraints
are well obeyed in general, for those LECs determined from
phenomenology.  The constraints on scattering lengths are discussed
in Ref.~\cite{penn, Distler} at $\cO(p^4)$. In Ref.~\cite{dita99prd}
the positivity constraints for the full amplitudes are discussed at
$\cO(p^6)$. Recently, these positivity constraints on the LECs are
carefully reinvestigated in Ref.~\cite{mateu}. In most of the
previous investigations, only the $\cO(p^4)$ lagrangian is examined,
because the large uncertainties existed for those $\cO(p^6)$
coefficients. On the other side, studies to the $\cO(p^6)$ LECs have
been recently extended~\cite{guo09prd}, comparing with the previous
estimation~\cite{bijnens97npb}. In order to have an understanding on
the positivity constraints in a more transparent way and to test the
newly determined $\cO(p^6)$ LECs, it is worthwhile to re-investigate
the positivity constraints at $\cO(p^6)$ level.
 We in this  note  firstly study
  in the leading order of $1/N_C$ expansion which enables us to
  obtain simple analytic expressions.
  We find that, at leading order of $1/N_C$, these constraints are
automatically satisfied, owing to the positivity of mass  and the
positivity of decay width of a resonance. We also investigate those
positivity constraints in the case of $N_C=3$, using the expressions
of $\cO(p^4)$ and $\cO(p^6)$ LECs derived in Ref.~\cite{guo09prd,ourmatch}
and find that they are well respected as well.

The $\pi\pi$ scattering amplitude is determined by the function
$A(s,t,u)$,
\begin{equation}
\begin{array}{l}
A\left[\pi^a(p_1)+\pi^b(p_2)\to \pi^c (p_3)+\pi^d(p_4)\right]
\,\, =\,\,
\\
\\
\quad  \delta^{ab}\delta^{cd} A(s,t,u)\,
+\, \delta^{ac}\delta^{bd} A(t,u,s)\,
+\, \delta^{ad}\delta^{bc} A(u,t,s)\, .
\end{array}
\end{equation}
We express the amplitude $A(s,t,u)$ explicitly in terms of LECs (
independent of the pseudo-Goldstone masses), momenta  and
pseudo-Goldstone masses:
\begin{eqnarray}
\label{op6ampop2} A(s,t,u)^{\rm \chi PT}\, = && \,
\frac{s-m_\pi^2}{f^2} +\frac{m_\pi^4}{f^4}\left( 8l_1 + 2l_3 \right)
- \frac{ 8 m_\pi^2s}{f^4}l_1 +\frac{s^2}{f^4}(2l_1+\frac{l_2}{2})
\nonumber
\\ &&
\, +\frac{(t-u)^2}{2f^4}l_2 -\frac{8 m_\pi^6}{f^6}l_3^2
+\frac{m_\pi^6}{f^6} \left(r_1+ 2 r_f\right) +\frac{m_\pi^4s}{f^6}
\left(r_2- 2 r_f\right) \nonumber
\\ &&
\, + \,  \frac{m_\pi^2s^2}{f^6}r_3 + \frac{m_\pi^2(t-u)^2}{f^6}r_4
+\frac{s^3}{f^6}r_5 +\frac{s(t-u)^2}{f^6}r_6 \,,
\end{eqnarray}
where $s=(p_1+p_2)^2,\, t=(p_1-p_3)^2,\,
u=(p_1-p_4)^2=4m_\pi^2-s-t$; $f$ is the pion decay constant in the
chiral limit and the chiral expansion of  the pion decay constant
$f_\pi$ up to $\cO(p^6)$ has been used,
\begin{equation} \label{fpi}
\hspace*{-0.25cm} f_\pi = f\left[1+\frac{l_4 m_\pi^2}{f^2} +(-2l_3
l_4+r_f)\frac{m_\pi^4}{f^4}+\cO(m_\pi^6)\right].
\end{equation}
In both expressions given above,
only  leading terms in the $1/N_C$ expansion are kept, comparing
with the original expression~\cite{bijnens97npb}. Comparing with the
analysis in Ref.~\cite{dita99prd}, instead of using the $b_i$
parameters~\cite{dita99prd,bijnens97npb}, which are combinations of
the $\cO{(p^4)}$ and $\cO(p^6)$ LECs, we have reexpressed the
amplitudes explicitly with the $\cO{(p^4)}$ LECs $l_i$ and the
$\cO(p^6)$ LECs $r_i$, which can make the analysis order by order
within the chiral expansion in a more transparent
way~\cite{ourmatch}.

There are three positivity constraints on the full $\pi\pi$
scattering amplitudes obtainable using forward dispersion relations:
\bqa
\frac{d^2}{d s^2 } T(\pi^0\pi^0\to\pi^0\pi^0)[s,t] >0 \,, \\
\frac{d^2}{d s^2 } T(\pi^+\pi^0\to\pi^+\pi^0)[s,t] >0 \,, \\
\frac{d^2}{d s^2 } T(\pi^+\pi^+\to\pi^+\pi^+)[s,t] >0 \,, \eqa which
are valid in a region  of the Mandelstam plane defined by $0 \leq t
\leq 4m_\pi^2$, $s \leq 4m_\pi^2$, $s + t \geq 0$ (herewith called
as extended Mandelstam triangle)~\cite{mateu, dita99prd}. Notice
that this region is larger than the conventional Mandelstam triangle
defined by $0\leq s,t,u<4m_\pi^2$. These inequalities lead
respectively to the following positivity constraints on the LECs:
 \bqa\label{fullcons1}
(l_1+l_2)+\frac{m_\pi^2}{2f^2}(r_3+3r_4+6r_5-2r_6)-\frac{3 t}{4 f^2}(r_5-3r_6) >0\,, \\
\label{fullcons2} l_2+\frac{2m_\pi^2}{f^2} r_4+ \frac{2t}{ f^2} r_6
>0 \,,
\\ \label{fullcons3}
(2l_1+3l_2)+\frac{m_\pi^2}{f^2} (r_3+5 r_4+12 r_5+4
r_6)-\frac{3s}{f^2} ( r_5+r_6) -\frac{t}{ f^2} (3r_5-5r_6) >0 \,.
\eqa
 It is known that, at level of ${\cal O}(p^4)$, there are only two
 independent constraints in the large $N_C$ limit. However, as seen
 from above equations, the three constraints are not degenerate in
 general. It is easy to check that merely in the special case when
 $2s+t=4m_\pi^2$ only two of the three constraints given above are
 independent.

Positivity constraints are also obtainable in a simpler way by
applying  optical theorem to forward dispersion relations, which
corresponds to a special case of taking $t=0$ in the above analysis.
In this way one  gets,
 \bqa\label{fullcons1t0}
(l_1+l_2)+\frac{m_\pi^2}{2f^2}(r_3+3r_4+6r_5-2r_6)>0\,, \\
\label{fullcons2t0} l_2+\frac{2m_\pi^2}{f^2} r_4 >0 \,,
\\\label{fullcons3t0} (2l_1+3l_2)+\frac{m_\pi^2}{f^2} (r_3+5 r_4+12
r_5+4 r_6)-\frac{3s}{ f^2} ( r_5+r_6)>0 \,. \eqa
 Notice that the first two equations in above are the $\cO(p^6)$ extensions of those obtained
 in Ref.~\cite{pham}.

Taking $t=4 m_\pi^2$, which is used in \cite{mateu},   leads to the
following results:
 \bqa\label{fullcons1t4}
(l_1+l_2)+\frac{m_\pi^2}{2f^2}(r_3+3r_4+16r_6)>0\,,
\\\label{fullcons2t4} l_2+\frac{2m_\pi^2}{f^2} (r_4+4r_6) >0 \,,\\
\label{fullcons3t4} (2l_1+3l_2)+\frac{m_\pi^2}{f^2} (r_3+5 r_4+24
r_6)-\frac{3s}{f^2} ( r_5+r_6)>0 \,. \eqa

Since the factors $m_\pi^2/2f^2$ and $m_\pi^2/4f^2$ are numerically
of $\cO(1)$, one has to verify whether the $\cO(p^6)$ LECs $r_i$
play an important role numerically in  above $\cO(p^6)$ relations.
Before making numerical analysis we notice that the above relations
can be rewritten in another form. In Ref.~\cite{ourmatch}, using the
partial wave dispersion relations and large $N_C$ technique, the
LECs can be reexpressed in terms of mass and decay width of
resonances without relying on any explicit resonance lagrangian:
\bqa\label{li} l_1 \,&=&\, \frac{16 \pi f^4}{3} \left(
\frac{\overline{\Gamma}_S}{ \overline{M}_S^5} \, -\,
\frac{ 9 \overline{\Gamma}_V}{\overline{M}_V^5} \right) \, ,\nonumber \\
 l_2 \, &=&\,  48\pi f^4 \frac{\overline{\Gamma}_V}{\overline{M}_V^5} \,,
\eqa
\bqa\label{ri}
    r_2-2r_f &=& \frac{64\pi f^6 \overline{\Gamma}_S}{\overline{M}_S^7} \left( 1+
\frac{\beta_{\rm S}}{3} +\frac{\gamma_{\rm S}}{6}   \right)+\frac{\pi f^6
\overline{\Gamma}_V}{\overline{M}_V^7} \left( 7584 + 1248 \beta_{\rm
V} + 144 \gamma_{\rm V}       \right) \,, \nonumber \\
r_3 &=&\frac{64 \pi f^6 \overline{\Gamma}_S}{3\overline{M}_S^7} \left(
1+\frac{\beta_{\rm S}}{2}\right) \, - \, \frac{768 \pi f^6
\overline{\Gamma}_V}{\overline{M}_V^7} ( 1 + \frac{3  \beta_{\rm
V}}{32} )  \,\,\, , \nonumber \\
 r_4&=&\frac{192 \pi f^6 \overline{\Gamma}_V}{\overline{M}_V^7}
 \left( 1 +\frac{\beta_{\rm V}}{8}\right) \,,
\nonumber \\
r_5&=&\frac{32 \pi f^6 \overline{\Gamma}_S}{3\overline{M}_S^7}+\frac{36 \pi f^6 \overline{\Gamma}_V}{\overline{M}_V^7}
\, ,\nonumber \\
r_6&=&\frac{12 \pi f^6 \overline{\Gamma}_V}{\overline{M}_V^7} \,  ,
\eqa
 where  subscripts $V$ and $S$ denote vector and scalar resonances, respectively;
$\overline{\Gamma}_R$ and $\overline{M}_R$ stand, respectively, for
the value of the $R$ resonance's width and mass in the chiral limit;
the $\cO(m_\pi^2)$ corrections are reflected in coefficients
$\alpha, \beta, \gamma$, which are defined as
\begin{equation}
\label{beta}
\frac{\Gamma_R}{M_R^5}=\frac{\overline{\Gamma}_R}{\overline{M}_R^{5}}
\left[1+
\beta_R\frac{m_\pi^2}{\overline{M}_R^{2}}+\cO(m_\pi^4)\right], \ee
 \be
\label{alphagamma}
\frac{\Gamma_R}{M_R^3}=\frac{\overline{\Gamma}_R}{\overline{M}_R^{3}} \left[1+
\alpha_R\frac{m_\pi^2}{\overline{M}_R^{2}}+\gamma_R\frac{m_\pi^4}{\overline{M}_R^{4}}+\cO(m_\pi^6)\right].
\end{equation}
Substituting Eqs.~(\ref{li}) and (\ref{ri})
into Eqs.~(\ref{fullcons1}--\ref{fullcons3}),
we can translate the positivity
constraints on LECs into the following simple form:
\begin{itemize}
\item from $\pi^0\pi^0 \rightarrow \pi^0\pi^0$,
\bqa\label{bspct0} { \overline{M}_S^2}+(\beta_S
+8){m_\pi^2}-\frac{3t}{2}> 0\,, \eqa
\item from $\pi^+\pi^0 \rightarrow \pi^+\pi^0$,
 \bqa\label{bvpct1}
{\overline{M}_V^2}+(\beta_V +8)m_\pi^2+\frac{t}{2}> 0 \,, \eqa
\item from $\pi^+\pi^+\rightarrow \pi^+\pi^+$,
\bqa\label{bsvpct2} \frac{(\beta_S +14)m_\pi^2-3s-3t+
\overline{M}_S^2}{9\overline{M}_S^7}\,\overline{\Gamma}_S \,+\,
\frac{(\beta_V +14)m_\pi^2-3s-t+
\overline{M}_V^2}{2\overline{M}_V^7}\,\overline{\Gamma}_V \,>0\,.
\eqa
\end{itemize}
One has the following observations from the above inequalities:
\begin{itemize}
\item In the leading order of chiral expansion, these positivity
constraints become automatic, owing to the positivity of mass and
width of resonances. This can also be clearly seen, for
example, by substituting Eq.~(\ref{li}) into
Eqs.~(\ref{fullcons1t0}) -- (\ref{fullcons3t0}).
\item Inside the Mandelstam triangle, these inequalities are even
automatically hold at $\cO(p^6)$ level. Notice that positivity of
the resonance width and mass, i.e., Eq.~(\ref{beta}), requires
$1+\beta_R\frac{m_\pi^2}{M_R^2}>0$. This condition enforces that the
three constraints Eqs.~(\ref{bspct0}) -- (\ref{bsvpct2}) are
unconditionally satisfied inside the Mandelstam triangle.
\item In the extended region of Mandelstam triangle,
Eqs.~(\ref{bspct0}) and (\ref{bvpct1}) are still automatically
satisfied, but Eq.~(\ref{bsvpct2}) is no longer the case. Set for
example $s=t=4m_\pi^2$, one gets,
$$\frac{(\beta_S
-10)m_\pi^2+
\overline{M}_S^2}{9\overline{M}_S^7}\,\overline{\Gamma}_S \,+\,
\frac{(\beta_V -2)m_\pi^2+
\overline{M}_V^2}{2\overline{M}_V^7}\,\overline{\Gamma}_V \,>0\,.
$$
Therefore this analysis shows that to discuss the positivity
condition in the enlarged region is useful in the sense that it
indeed provides stronger constraints.
 Nevertheless, from the values given in Ref.~\cite{guo09prd}, i.e.,
 $\beta_S=2 \pm 8$ and $\beta_V=-7.7\pm 0.3$,
  $\overline{\Gamma}_V= 177.8\pm 2.5$~MeV, $\overline{\Gamma}_S= 600\pm 300$~MeV,
$\overline{M}_V=764.3\pm 1.1$~MeV and $\overline{M}_S=980\pm
40$~MeV, that the Eq.~(\ref{bsvpct2}) is still satisfied very well
numerically.
\item  In Ref.~\cite{mateu}, ${\cal O}(p^4)$ amplitudes with chiral loops
 are analyzed and it is concluded that the most stringent bounds are
 always found at $t=4m_\pi^2$. In the $\cO(p^6)$ case at the leading order
 of $1/N_C$ expansion, the situation can be
 somewhat different. For example, by taking $t=4 m_\pi^2$ for
 Eqs.~(\ref{bspct0}) and(\ref{bsvpct2}), ones finds out that the
 constraints  are stronger than taking $t=0$, but for Eq.~(\ref{bvpct1})
 is instead weaker.
\end{itemize}

Positivity constraint on partial waves are also discussed in the
literature~\cite{dispersive}.  The $D$ wave projection of the
$\pi^0\pi^0 \to \pi^0\pi^0$ amplitude is
 \bqa && T^{D}(\pi^0\pi^0 \to
\pi^0\pi^0)= \frac{(s-4m_\pi^2)^2}{1920\pi f^6}\times \nonumber \\
&& \times \bigg[ 4 f^2(l_1+l_2)+ 2
m_\pi^2(r_3+3r_4+6r_5-2r_6)-3(r_5-3r_6)s \bigg]\,.  \eqa Positivity
requirement leads to \bqa
(l_1+l_2)+\frac{m_\pi^2}{2f^2}(r_3+3r_4+6r_5-2r_6)-\frac{3 s}{4
f^2}(r_5-3r_6) >0\,. \eqa We find the constraint from the $D$ wave
amplitude of $\pi^0\pi^0 \to \pi^0\pi^0$ is the same as the one from
the full amplitude constraint given in Eq.~(\ref{fullcons1}).

In above analysis, we have made it clear that the positivity
constraints are well satisfied in the large $N_C$ limit, and in most
cases they are automatically obeyed, especially inside  the region
of Mandelstam triangle. The reason behind may be explained as, when
chiral perturbation theory is embedded into resonance chiral theory,
it has a genuine high energy behavior. The possible factors that
could lead to the violation of positivity conditions as emphasized
in Ref.~\cite{dispersive} are hence no longer worrisome.

The above analysis are confined to the case of leading order of
$1/N_C$ expansion. In the following, the effect of the $1/N_C$
corrections will be discussed. Discussions at $\cO(p^6)$ have been
partly made in Ref.~\cite{mateu,dita99prd} and the conclusion is
that the constraints are in general very well satisfied for
realistic value of LECs~\cite{dita99prd}. Here we will take another
point of view to check the effect of the sub-leading order of
$1/N_C$. In the large $N_C$ limit although one can predict the
$\chi$PT LECs by integrating out the heavy resonances, such as the
expressions given in Eqs.~(\ref{li}) and (\ref{ri}) or the ones in
Ref.~\cite{lec}, it is however not clear at which scale these
expressions apply. The scale dependence of the $\chi$PT LECs is of
higher order effect in $1/N_C$ expansion. At $\cO(p^4)$ level, it is
demonstrated the resonance saturation works pretty well at the scale
of $\mu= M_\rho$~\cite{lec}. However there is no strict proof
that the resonance saturation must happen exactly at the scale of
the resonance mass $M_R$. Instead of making the constraints on the
explicit value of the LECs~\cite{mateu,dita99prd}, we will
investigate the positivity constraints on the saturation scale $\mu$
by taking into account the loop contributions given
in~\cite{lec}. We assume the renormalized LECs $l_i^r$ and
$r_i^r$ in the $\pi\pi$ scattering amplitudes~\cite{bijnens97npb}
are provided by Eqs.(\ref{li}) and (\ref{ri}). For $l_3^r, l_4^r$,
we use the results from Ref.~\cite{lec} \bqa
l_3^r=4\frac{c_m(c_m-c_d)}{\overline{M}_S^2}\,, \qquad
l_4^r=4\frac{c_m c_d}{\overline{M}_S^2}\,, \eqa where the values of
$c_d=( 26 \pm 7) {\rm MeV}, c_m= (80 \pm 21) {\rm MeV}$ will be taken from Ref.~\cite{guo09prd}.

\begin{figure}[h]%
\begin{center}%
\mbox{\epsfxsize=75mm\epsffile{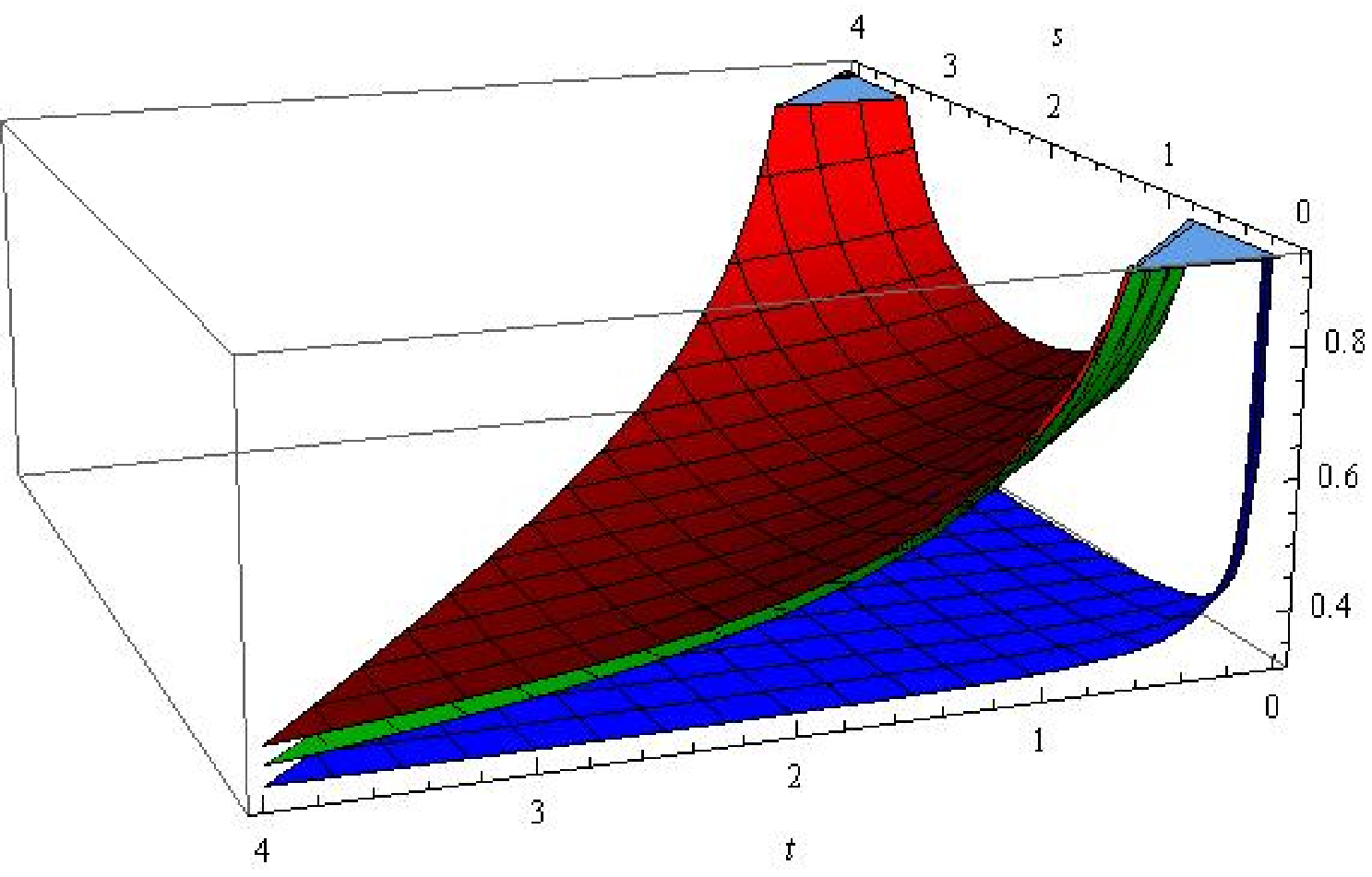}}%
\mbox{\epsfxsize=75mm\epsffile{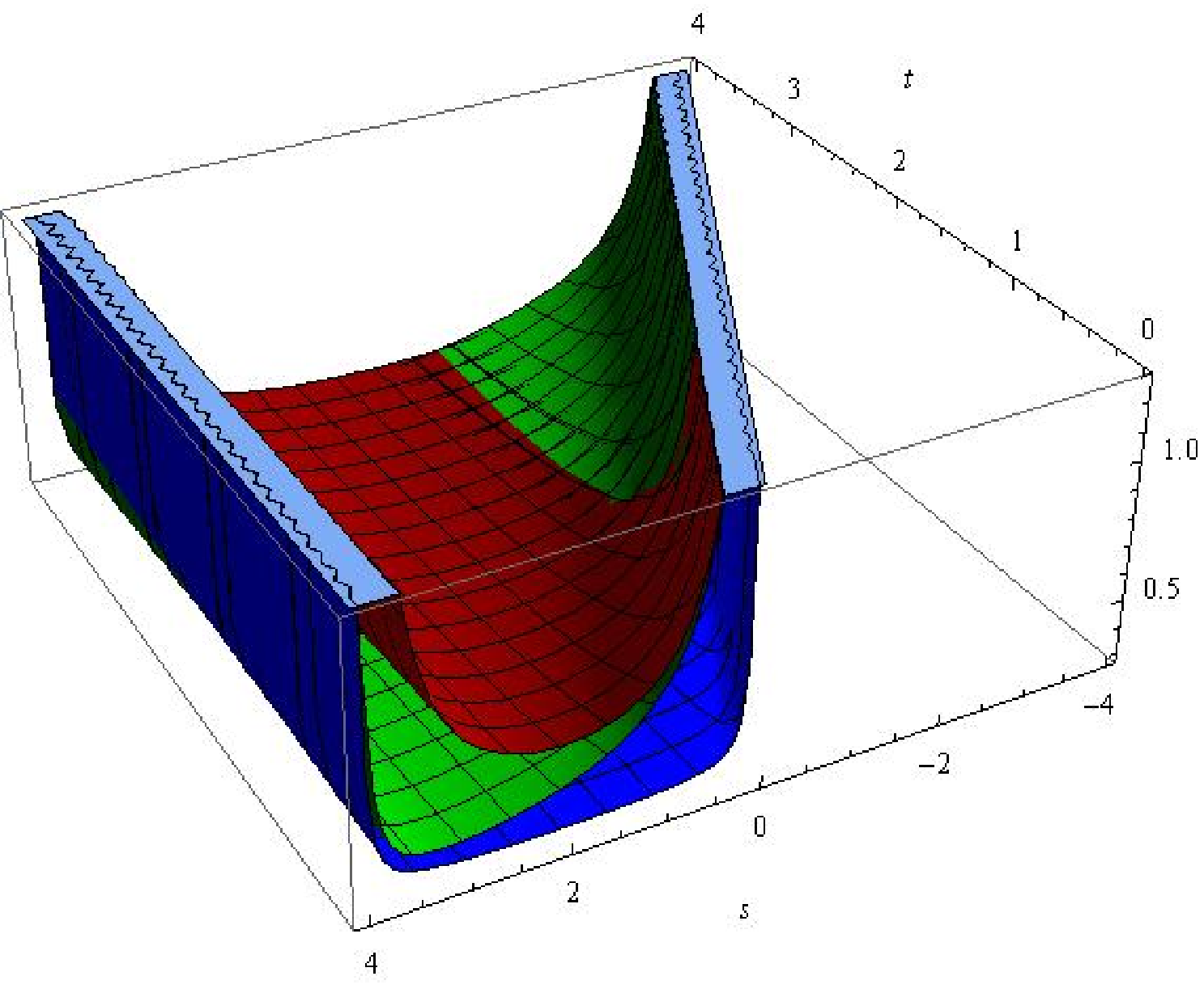}}%
\caption{\label{positivity} Left: scattering amplitudes in the
Mandelstam triangle; right: in the extended region. The red curve
corresponds to the case $\pi^0\pi^0\to \pi^0\pi^0$, green one
corresponds to $\pi^+\pi^+\to\pi^+\pi^+$, blue  one corresponds to,
$\pi^+\pi^0\to\pi^+\pi^0$, respectively. Scale $\mu=770$MeV. The amplitudes are given in unit of $m_\pi^4$ and $s,t$ are
given in unit of $m_\pi^2$.}
\end{center}%
\end{figure}

Since we have fixed the renormalized LECs at an unknown scale $\mu$,
the $\pi\pi$ scattering amplitudes given in~\cite{bijnens97npb} will
be explicitly dependent on $\mu$. In this way, the positivity constraints
on the $\pi\pi$ scattering amplitudes are translated into the
constraints on the saturation scale $\mu$. We find the positivity
constraints are all well satisfied at $\mu=770$ MeV for the three
channels within the Mandelstam triangle, and inside the extended
region as well.
The most stringent constraint on $\mu$ we find
appears in $\pi^0\pi^0 \to \pi^0\pi^0$ channel at $s=0, t=4m_\pi^2$:
$ \mu \gtrsim 245 {\rm MeV}$. The reason behind can be explained
as that the  $\pi^0\pi^0 \to \pi^0\pi^0$ process is only contributed by
the scalar resonances and indeed scalar resonances get significant
contribution  from the sub-leading order of $1/N_C$ expansion.
Since the value of $ \mu = 245 {\rm MeV}$ seems to be too small
to be realistic, one can safely conclude that the positivity
constraints are indeed very well satisfied in reality,  at
$\cO(p^6)$ level. In Fig.~\ref{positivity}, we plot the value of
various  amplitudes in the Mandelstam triangle and in the extended
region, for $\mu=770$MeV. As it has already been mentioned in~\cite{mateu}
that the scalar one loop two point function is not smooth at threshold,
we also find the uneven behavior of the amplitudes near the thresholds
in Fig.~\ref{positivity}.

The violation of positivity constraints signals the break down of
effective theory. In this note, we extend the previous study on
positivity constraints  to $\cO(p^6)$ and find the current
determination of $l_i$~\cite{fitli} and $r_i$~\cite{guo09prd,ourmatch} well
satisfies the positivity relations given in the large $N_C$ limit,
and also in reality.

This work is supported in part by the European Commission (EC) RTN
network, Contract No.  MRTN-CT-2006-035482  (FLAVIAnet), and also by
National Nature Science Foundation of China under Contract
Nos. 10875001, 
10721063. 

\end{document}